\documentstyle[12pt,epsfig]{article}
\newcommand{\Frac}[2]%
{{\textstyle \frac{\mbox{\footnotesize $#1$}\rule[-0.9mm]{0mm}{1mm}}%
{\mbox{\footnotesize $#2$}\rule{0mm}{3.1mm}}}}
\renewcommand{\thefootnote}{\fnsymbol{footnote}}

\newcommand{\footfrac}[2]%

\begin{document}
\begin{titlepage}
\vspace*{-12 mm}
\noindent
\begin{flushright}
\begin{tabular}{l@{}}
\end{tabular}
\end{flushright}
\vskip 12 mm
\begin{center}
{\large \bf
The proton spin puzzle: where are we today ?
}
\\[14 mm]
{\bf Steven D. Bass}
\\[10mm]
{\em
Institute for Theoretical Physics, \\
Universit\"at Innsbruck,
Technikerstrasse 25, Innsbruck, A 6020 Austria
\\[5mm]
}
\vspace{0.5cm}

\end{center}
\vskip 20 mm
\begin{abstract}
\noindent 
The proton spin puzzle has challenged our understanding of QCD
for the last 20 years.
New measurements of 
polarized glue, 
valence and sea quark polarization, 
including strange quark polarization, are available.
What is new and exciting in the data, and what might this tell us about
the structure of the proton ?
The proton spin puzzle seems to be telling us about 
the
interplay of valence quarks with the complex vacuum structure of QCD.

\end{abstract}

\vspace{9.0cm}

\end{titlepage}
\renewcommand{\labelenumi}{(\alph{enumi})}
\renewcommand{\labelenumii}{(\roman{enumii})}
\renewcommand{\thefootnote}{\arabic{footnote}}

\newpage
\baselineskip=6truemm

\section{Introduction}

Protons behave like spinning tops. 
Unlike classical tops, however, the spin of these particles is an 
intrinsic quantum mechanical phenomenon. 
This spin is responsible for many fundamental properties of matter, 
including the proton's magnetic moment, 
the different phases of matter in low-temperature physics, 
the properties of neutron stars, 
and the stability of the known universe. 
How is the proton's spin built up from its quark and gluon constituents ?

It is 20 years since the European Muon Collaboration (EMC)
published their polarized deep inelastic measurement of the proton's
$g_1$ spin dependent structure function
and
the flavour-singlet axial-charge $g_A^{(0)}|_{\rm pDIS}$
\cite{emc}.
Their results suggested that the quarks' intrinsic spin
contributes little of the proton's spin.
The challenge to understand the spin structure of the proton
\cite{bassrmp,bassbook}
has inspired a vast programme of theoretical activity and new experiments at
CERN, DESY, JLab, RHIC and SLAC.
Where are we today ?

We start by recalling the $g_1$ spin sum-rules.

These are derived starting from the dispersion relation for polarized
photon-nucleon scattering and, for deep inelastic scattering,
the light-cone operator product expansion.
One finds that the first moment of  the $g_1$ structure function
is related
to the scale-invariant axial charges of the target nucleon by
\begin{eqnarray}
\int_0^1 dx \ g_1^p (x,Q^2)
&=&
\Biggl( {1 \over 12} g_A^{(3)} + {1 \over 36} g_A^{(8)} \Biggr)
\Bigl\{1 + \sum_{\ell\geq 1} c_{{\rm NS} \ell\,}
\alpha_s^{\ell}(Q)\Bigr\}
\nonumber \\
& &
+ {1 \over 9} g_A^{(0)}|_{\rm inv}
\Bigl\{1 + \sum_{\ell\geq 1} c_{{\rm S} \ell\,}
\alpha_s^{\ell}(Q)\Bigr\}  +  {\cal O}({1 \over Q^2})
 + \ \beta_{\infty}
.
\nonumber \\
\label{eqc50}
\end{eqnarray}
Here $g_A^{(3)}$, $g_A^{(8)}$ and $g_A^{(0)}|_{\rm inv}$ are the
isovector, SU(3) octet and scale-invariant  flavour-singlet axial
charges respectively. The flavour non-singlet $c_{{\rm NS} \ell}$
and singlet $c_{{\rm S} \ell}$ Wilson coefficients are calculable in
$\ell$-loop perturbative QCD \cite{Larin:1997}. 
The term $\beta_{\infty}$
represents a possible leading-twist subtraction constant
from the circle at infinity when one closes the contour in the
complex plane in the dispersion relation \cite{bassrmp}.
If finite, the subtraction constant affects just the first moment
sum-rule and, thus, corresponds to Bjorken $x=0$.
The first
moment of $g_1$ plus the subtraction constant, if finite, is equal
to the axial-charge contribution. The subtraction constant
corresponds to a real term in the spin-dependent part of the forward
Compton amplitude.

In terms of the flavour dependent axial-charges
\begin{equation}
2M s_{\mu} \Delta q =
\langle p,s |
{\overline q} \gamma_{\mu} \gamma_5 q
| p,s \rangle
\label{eqc55}
\end{equation}
the isovector, octet and singlet axial charges are:
\begin{eqnarray}
g_A^{(3)} &=& \Delta u - \Delta d
\nonumber \\
g_A^{(8)} &=& \Delta u + \Delta d - 2 \Delta s
\nonumber \\
g_A^{(0)}|_{\rm inv}/E(\alpha_s) 
\equiv 
g_A^{(0)} 
&=& \Delta u + \Delta d + \Delta s
.
\label{eqc56}
\end{eqnarray}
Here
\begin{equation}
E(\alpha_s) = \exp \int^{\alpha_s}_0 \! d{\tilde \alpha_s}\,
\gamma({\tilde \alpha_s})/\beta({\tilde \alpha_s})
\label{eqc54}
\end{equation}
is a renormalization group factor
which corrects
for the (two loop) non-zero anomalous dimension
$\gamma(\alpha_s)$
of the singlet axial-vector current 
\begin{equation}
J_{\mu5} = 
\bar{u}\gamma_\mu\gamma_5u
                  + \bar{d}\gamma_\mu\gamma_5d
                  + \bar{s}\gamma_\mu\gamma_5s 
\label{eqc53}
\end{equation}
which goes to one in the limit $Q^2 \rightarrow \infty$;
$\beta (\alpha_s)$ is the QCD beta function.
We are free to choose
the QCD coupling $\alpha_s(\mu)$ at either a hard or a soft scale
$\mu$.
The singlet axial charge $g_A^{(0)}|_{\rm inv}$
is independent of the renormalization scale $\mu$
and corresponds
to 
$g_A^{(0)}(Q^2)$ evaluated in the limit $Q^2 \rightarrow \infty$.
The perturbative QCD expansion of $E(\alpha_s)$ 
remains close to one -- even for large values of $\alpha_s$.
If we take $\alpha_s \sim 0.6$ as typical of the infra-red then
$
E(\alpha_s) \simeq
1 - 0.13 - 0.03 + ... = 0.84 + ...
$
where -0.13 and -0.03
are the ${\cal O}(\alpha_s)$ and ${\cal O}(\alpha_s^2)$
corrections respectively.

If one assumes no twist-two subtraction constant 
($\beta_{\infty} = O(1/Q^2)$)
then the axial charge contributions saturate the first moment
at leading twist.
The isovector axial-charge is measured independently in neutron
beta-decays
($g_A^{(3)} = 1.270 \pm 0.003$ \cite{PDG:2004})
and the octet axial charge is commonly taken 
to be the value extracted 
from hyperon beta-decays assuming good SU(3) properties
($g_A^{(8)} = 0.58 \pm 0.03$ \cite{fec}).
From the first moment of $g_1$,
polarized deep inelastic scattering experiments have been
interpreted
to imply a small value for the flavour-singlet axial-charge.
Inclusive $g_1$ data with $Q^2 > 1$ GeV$^2$
give \cite{compassnlo}
\begin{equation}
g_A^{(0)}|_{\rm pDIS, Q^2 \rightarrow \infty}
=
0.33 \pm 0.03 ({\rm stat.}) \pm 0.05 ({\rm syst.})
\end{equation}
-- considerably less than the value of $g_A^{(8)}$
quoted above. 
In the naive
parton model $g_A^{(0)}|_{\rm pDIS}$ is interpreted as the fraction
of the proton's spin which is carried by the intrinsic spin of its
quark and antiquark constituents. 
When combined with 
$g_A^{(8)} = 0.58 \pm 0.03$ 
this value corresponds to a negative
strange-quark polarization
\begin{equation}
\Delta s_{Q^2 \rightarrow \infty}
=
{1 \over 3}
(g_A^{(0)}|_{\rm pDIS, Q^2 \rightarrow \infty} - g_A^{(8)})
=
- 0.08 \pm 0.01 ({\rm stat.}) \pm 0.02 ({\rm syst.})
\end{equation}
-- that is,
polarized in the opposite direction to the spin of the proton.
The corresponding up and down quark polarizations are likewise 
extracted to be
\begin{eqnarray}
\Delta u_{Q^2 \rightarrow \infty}
= 
0.84 \pm 0.01 ({\rm stat.}) \pm 0.02 ({\rm syst.})
\nonumber \\
\Delta d_{Q^2 \rightarrow \infty}
=
-0.43 \pm 0.01 ({\rm stat.}) \pm 0.02 ({\rm syst.})
\end{eqnarray}
Relativistic quark models generally predict values $g_A^{(0)} \sim 0.6$
with little polarized strangeness in the nucleon\cite{Ellis:1974}
in agreement with the value of $g_A^{(8)}$ extracted from SU(3).
The Bjorken sum-rule for the isovector part of $g_1$, 
$
\int_0^1 dx g_1^{p-n}
=
{1 \over 6} g_A^{(3)} 
\Bigl\{1 + \sum_{\ell\geq 1} c_{{\rm NS} \ell\,}
\alpha_s^{\ell}(Q)\Bigr\}
$,
has been confirmed in polarized deep inelastic scattering experiments at
the level of 10\% \cite{bjsr}.

The results from polarized deep inelastic scattering pose the following 
questions:
\begin{itemize}
\item
How is the spin ${1 \over 2}$ of the proton built up from the spin and
orbital angular momentum of the quarks and gluons inside ?
\item
Why is the quark spin content $g_A^{(0)}|_{\rm pDIS}$ so small ?
\item
How about $g_A^{(0)} \neq g_A^{(8)}$ ?
What separates the values of the octet and singlet axial-charges ?
\item
Is the proton spin puzzle a valence quark or sea/glue effect ?
\end{itemize}
We next discuss the experiments that have been performed to address
these questions.

\section{The shape of $g_1$}

\begin{figure}
\centerline{\psfig{file=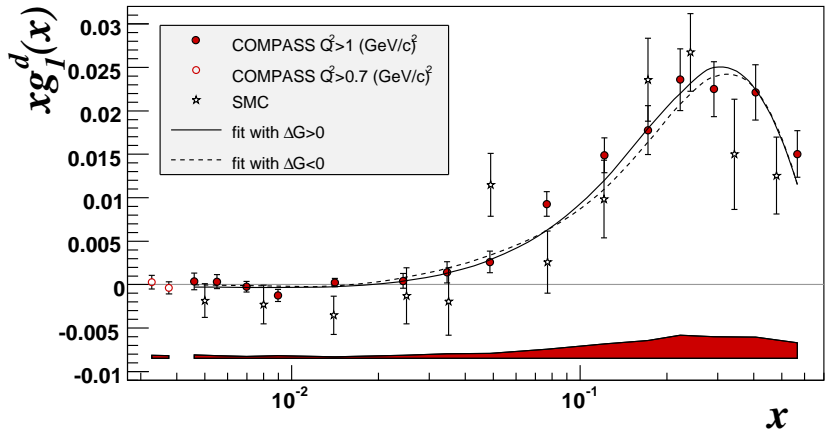,width=4.5in}}
\vspace*{5pt}
\caption{$g_1^d$ data from COMPASS $^7$.}
\label{fig:fig4}
\end{figure}

Deep inelastic measurements of $g_1$ have been performed in experiments at
CERN, DESY, JLab and SLAC.
There is a general consistency among all data sets.
COMPASS are yielding precise new data at small $x$, down to $x \sim 0.004$, 
which is shown for $g_1^d$ in Fig.1.
JLab are focussed on the large $x$ region. To test deep inelastic
sum-rules it is necessary to have all data points at the same value
of $Q^2$. In the experiments the different data points are measured
at different values of $Q^2$, viz. $x_{\rm expt.}(Q^2)$.
Next-to-leading order (NLO) QCD-motivated fits taking into account
the scaling violations associated with perturbative QCD are
frequently used to evolve all the data points to the same $Q^2$.

The COMPASS measurements of the deuteron spin structure function $g_1^d$ 
show the remarkable feature that $g_1^d$ 
is consistent with zero in the small $x$ region between 0.004 and 0.02
\cite{compassnlo}.
In contrast, the isovector part of $g_1$ is observed to rise at
small $x$ ($0.01 < x < 0.1$) as $\sim x^{-0.5}$ and is much bigger
than the isoscalar part of $g_1$. 
This is in
sharp contrast to the situation in the unpolarized structure
function $F_2$ where the small $x$ region is dominated by isoscalar
pomeron exchange. The evolution of the Bjorken integral
$\int_{x_{\rm min}}^1 dx g_1^{p - n}$ as a function of $x_{min}$ 
as well as the isosinglet integral
$\int_{x_{\rm min}}^1 dx g_1^{p + n}$
are
shown in Fig. 2 (HERMES data\cite{hermes}).
About 50\% of the Bjorken sum-rule 
$\int_0^1 dx g_1^{p-n}$
comes from $x$ values below about 0.12. 
The $g_1^{p-n}$ data are consistent with
quark model and perturbative QCD predictions in the valence region
$x > 0.2$ \cite{epja}. The size of $g_A^{(3)}$ forces us to accept a
large contribution from small $x$ and the observed rise in $g_1^{p -n}$ 
is required to fulfil this non-perturbative constraint,
perhaps signifying a hard Regge exchange\cite{bassmb} 
like the hard pomeron in unpolarized deep inelastic scattering\cite{pvl}.

The ``missing spin'' is associated with a ``collapse'' 
in the isosinglet part of $g_1$ to something close to 
zero instead of a valence-like rise 
$\sim x^{-0.5}$ for $x$ less than about 0.03.
The isosinglet integral appears to converge at $x_{\rm min} \sim 0.1$.
This isosinglet part is the sum of 
SU(3)-flavour singlet and octet contributions.
If there were a large positive polarized gluon contribution 
to the proton's spin, this would act to
drive the small $x$ part of the singlet part of $g_1$ negative\cite{bt91}
-- 
that is, acting in the opposite direction to any valence-like
rise at small $x$.
However, gluon polarization measurements at COMPASS, HERMES and RHIC
constrain this spin contribution to be small in measured kinematics
-- see below -- 
meaning that the sum of valence and sea quark contributions
is suppressed at small $x$.
(Soft Regge theory predicts that the singlet term should 
 behave as
 $\sim N \ln x$ in the small $x$ limit, 
 with the coefficient $N$ to be determined from experiment\cite{sbpvl,fec94}.)
Further data from HERMES and COMPASS 
involving semi-inclusive measurements of fast pions and kaons 
in the final state is being 
used to constrain the sea and valence
quark spin contributions and reveals no evidence for polarized
strangeness, anti-up or anti-down spin polarization in the proton
(in apparent contrast to the extraction of negative strangeness
 polarization extracted from inclusive measurements of $g_1$).
We next discuss this spin decomposition and the different measurements.

\begin{figure}
\centerline{\psfig{figure=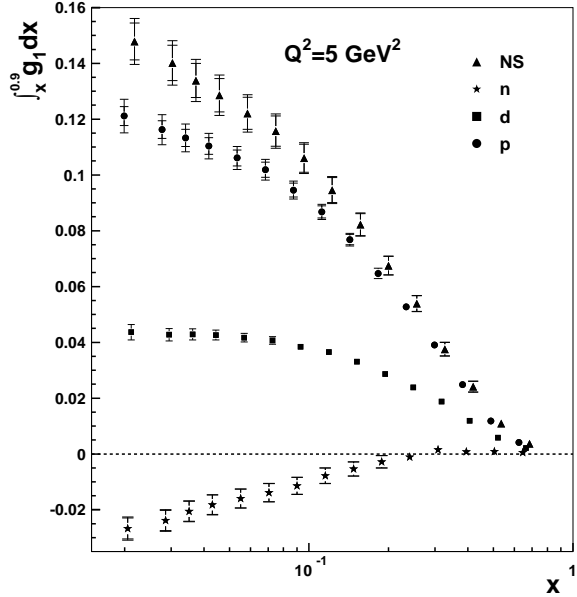,width=3.0in}}
\caption{Convergence of the first moment integrals for the proton,
 neutron, deuteron and isovector (NS) $g_1$ combination in HERMES
 data$^{10}$.}
\label{fig:fig4a}
\end{figure}

\section{Spin and the singlet axial charge}

There has been considerable theoretical effort to understand
the flavour-singlet axial-charge in QCD.
QCD theoretical analysis leads to the formula
\begin{equation}
g_A^{(0)}
=
\biggl(
\sum_q \Delta q - 3 {\alpha_s \over 2 \pi} \Delta g \biggr)_{\rm partons}
+ {\cal C}_{\infty}
.
\label{eqa10}
\end{equation}
Here $\Delta g_{\rm partons}$ is the amount of spin carried
by polarized
gluon partons in the polarized proton
($\alpha_s \Delta g \sim {\tt constant}$ as $Q^2 \rightarrow \infty$
\cite{ar,et})
and
$\Delta q_{\rm partons}$ measures the spin carried by quarks
and
antiquarks
carrying ``soft'' transverse momentum $k_t^2 \sim P^2, m^2$
where
$P$ is a typical gluon virtuality
and
$m$ is the light quark mass
\cite{ar,et,ccm,bint}.
The polarized gluon term is associated with events in polarized
deep inelastic scattering where the hard photon strikes a
quark or antiquark generated from photon-gluon fusion and
carrying $k_t^2 \sim Q^2$ \cite{ccm}.
${\cal C}_{\infty}$ denotes a potential non-perturbative gluon
topological contribution
\cite{topology}
which is associated with the possible subtraction constant in
the
dispersion relation for $g_1$ \cite{bassrmp}.
If finite it would mean that
$\lim_{\epsilon \rightarrow 0} \int_{\epsilon}^1 dx g_1$
will measure
the difference of
the singlet axial-charge and the subtraction constant contribution;
that is, polarized deep inelastic scattering measures the combination
$g_A^{(0)}|_{\rm pDIS} = g_A^{(0)} - C_{\infty}$.

Possible explanations for the small value of $g_A^{(0)}|_{\rm pDIS}$
extracted
from the polarized deep inelastic experiments
include
screening from positive gluon polarization,
negative strangeness polarization in the nucleon,
a subtraction at infinity in the dispersion relation for $g_1$
associated with non-perturbative gluon topology 
and
connections to axial U(1) dynamics \cite{tgv,shore,hf,bass99}.

One would like to understand the dynamics which appears to suppress 
the singlet axial-charge extracted from polarized deep inelastic
scattering relative to the OZI prediction $g_A^{(0)} = g_A^{(8)}
\sim 0.6$ and also the sum-rule for the longitudinal spin structure
of the nucleon
\begin{equation}
{1 \over 2} = {1 \over 2} \sum_q \Delta q + \Delta g + L_q + L_g
\end{equation}
where $L_q$ and $L_g$ denote the orbital angular momentum contributions.

There is presently a vigorous programme to disentangle the different
contributions. Key experiments involve semi-inclusive polarized deep
inelastic scattering (COMPASS and HERMES) and polarized
proton-proton collisions (PHENIX and STAR at RHIC).

\subsection{NLO QCD motivated fits to $g_1$}

The first attempts to extract information about gluon polarization
in the polarized nucleon used next-to-leading order (NLO)
QCD-motivated fits to inclusive $g_1$ data.

Similar to the analysis that is carried out on unpolarized data, global
NLO perturbative QCD analyses
have been performed on the polarized structure function data sets.
The aim is to extract the polarized quark and gluon parton distributions.
These QCD fits are performed within a given factorization scheme.
New fits are now being produced taking into account all
the available
data including new data from polarized semi-inclusive
deep inelastic scattering.
The largest uncertainties in these fits are associated with 
the ansatz
chosen for the shape of the spin-dependent quark and gluon
distributions at a given input scale.
Further, the SU(3) value of $g_A^{(8)}$ ($=0.58 \pm 0.03$) 
is assumed in these fits.
Fits to the most recent world data on $g_1$ give ``small'' values of 
$\Delta g$:
$|\Delta g|\simeq 0.2 - 0.3$ 
for $Q^2=3$~GeV$^2$ \cite{compassnlo,leader}.
To go further more direct measurements involving glue sensitive
observables are needed to really extract the magnitude of
$\Delta g$ and
the shape of $\Delta g (x, Q^2)$
including any possible nodes in the distribution function.

\subsection{Gluon polarization}

\begin{figure}[t!]
\begin{center}
\epsfig{file=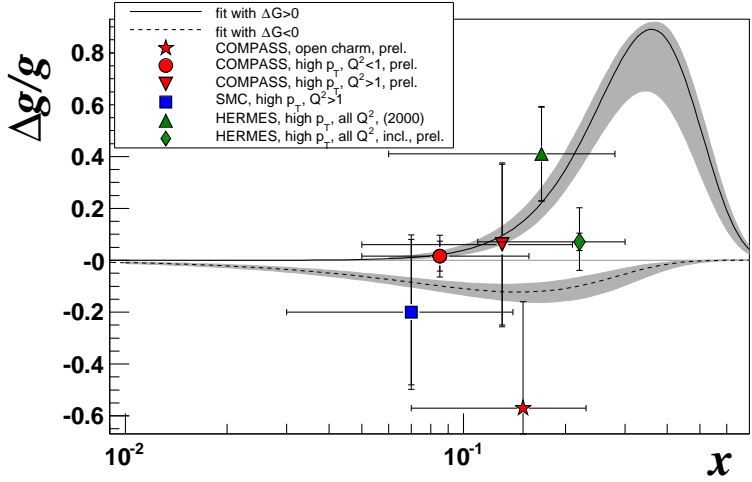
,width=0.7\textwidth}
\end{center}
\caption{Measurements of gluon polarization from COMPASS, HERMES and SMC
together with the COMPASS fits 
for $\Delta g(x)/g(x)$ at $Q^2$ = 3 GeV$^2$ 
corresponding to $\Delta g > 0$ and $\Delta g < 0$ $^{32}$.}
\label{fig:Delta_GbyG}
\end{figure}

There is a vigorous and ambitious global programme to measure
$\Delta g$. Interesting channels include gluon mediated processes in
semi-inclusive polarized deep inelastic scattering (COMPASS) and
hard QCD processes in high energy polarized proton-proton collisions
at RHIC.

The first experimental attempt to look at gluon polarization was made
by the FNAL E581/704 Collaboration which measured 
the double-spin asymmetry ${\cal A}_{LL}$ 
for inclusive multi-$\gamma$ and $\pi^0 \pi^0$ production 
with a 200 GeV polarized proton beam and a polarized proton target
suggesting that $\Delta g / g$ is not so large in the region of
$0.05 < x_g < 0.35$ \cite{Adams:1994}.

\begin{table}[b!]
\caption{Polarized gluon measurements from deep inelastic experiments.}
\vspace{3ex}
{\begin{tabular}{@{}lllrr@{}} 
  Experiment  &  process            &  $\langle x_g \rangle$
& $\langle \mu^2 \rangle$ (GeV$^2$) &  $\Delta g / g$   \\
\hline
 
HERMES      &  hadron pairs        & 0.17 &  $ \sim 2 $  &
$ 0.41 \pm 0.18 \pm 0.03$
\\
HERMES      &  inclusive hadrons   & 0.22 &  $ 1.35 $    &
$ 0.071 \pm 0.034 ^{+0.105}_{-0.127}$
\\
SMC         &  hadron pairs        & 0.07 &              &
$ -0.20 \pm 0.28 \pm 0.10$
\\
COMPASS     & hadron pairs, $Q^2 < 1$  &  $ 0.085 $  &  $ \sim 3 $ &
$0.016 \pm 0.058 \pm 0.054$ 
\\
COMPASS & hadron pairs, $Q^2 > 1$  & $0.082 $ &  $\sim 3$          &
$0.08 \pm 0.10 \pm 0.05$ 
\\
COMPASS & open charm & $0.11 $ &                          13       &
$-0.49 \pm 0.27 \pm 0.11 $ \\
\end{tabular}
}
\label{tab:table2}
\end{table}

COMPASS has been conceived to measure $\Delta g$ via the study of
the photon-gluon fusion process. 
The cross-section for this process is directly related 
to the gluon density at the Born level. 
The experimental technique consists of the
reconstruction of charmed mesons \cite{compasscharm}
or 
high $p_t$ particles in the final state \cite{compassdeltag}
to access $\Delta g$. 
The high $p_t$ particles method leads to samples with
larger statistics but these have larger background contributions
from QCD Compton processes and fragmentation. High $p_t$ charged
particle production has been used in earlier attempts by HERMES
\cite{hermesdeltag} and SMC \cite{smcdeltag} to access gluon
polarization. 
These measurements are listed in Table 1 and shown in Fig. 3 
for $x_g \sim 0.1$.

\begin{figure}[b!]
\includegraphics{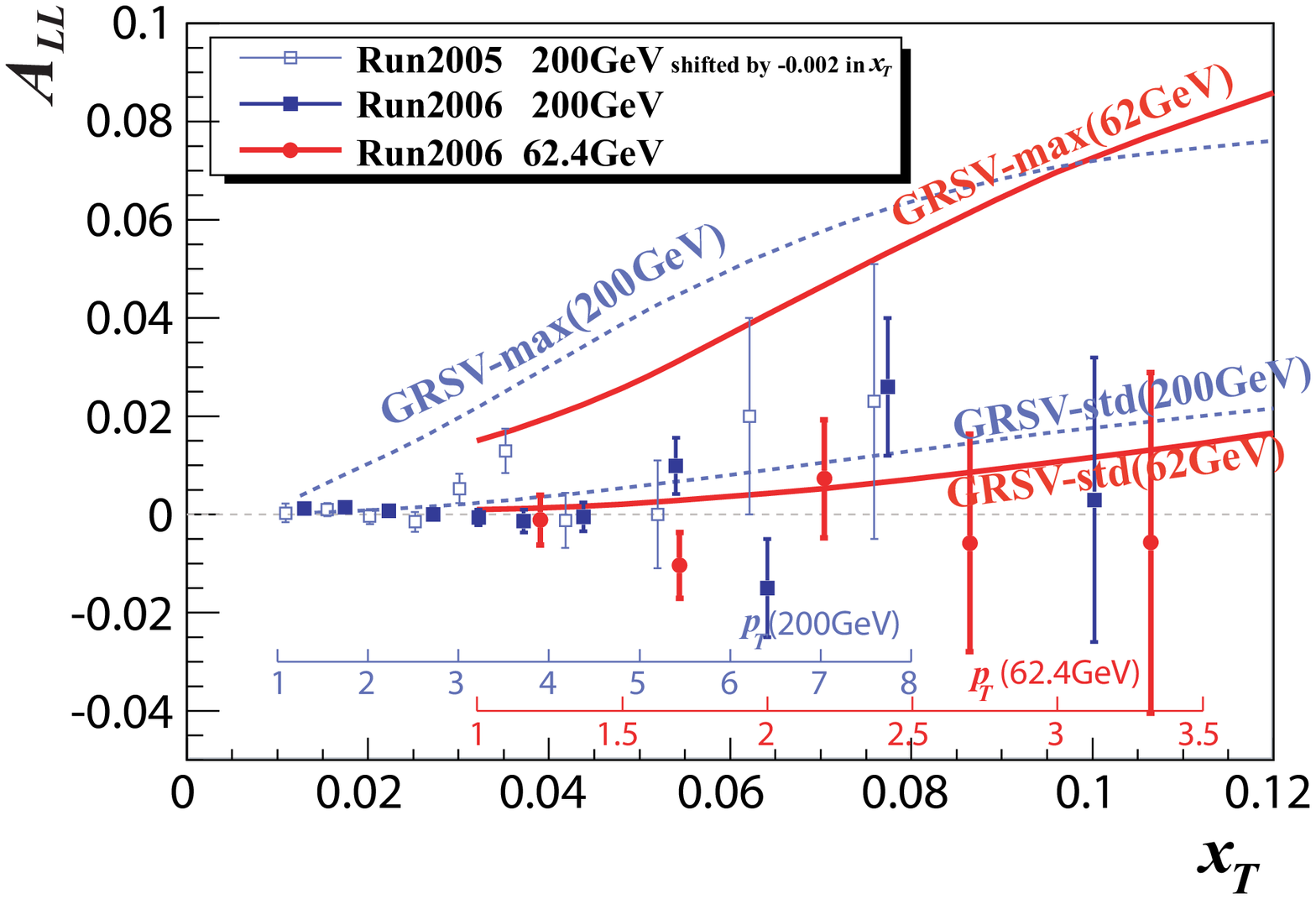}
\includegraphics{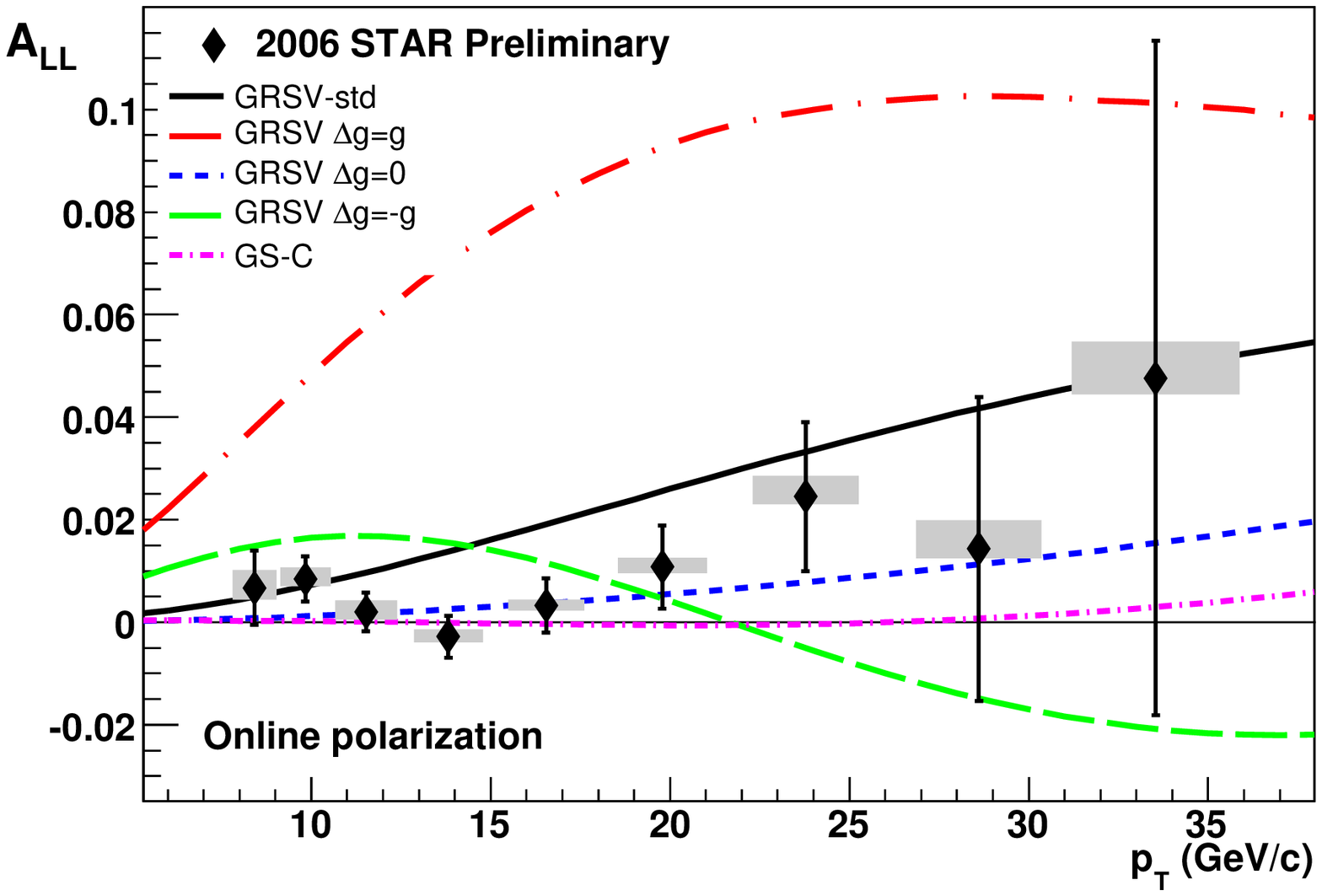}
\vspace{12.0cm}
{\caption[Delta]
{PHENIX results on ${\cal A}_{LL}^{\pi^0}$
 together
 with the predictions from various QCD fits at $s^{1 \over 2} = 200$ GeV 
 \cite{phenix200}
 and 62.4 GeV \cite{phenix62}, where $x_T = 2 p_T / \sqrt{s}$ (above).
 STAR data on the longitudinal double spin inclusive jet
 asymmetry $A_{LL}$ at $\sqrt{s} = 200$ GeV versus jet $p_T$
 \cite{star2008} (below).
}
\label{fig:fig8}}
\end{figure}

The hunt for $\Delta g$ is one of the main physics drives for polarized RHIC. 
Experiments using the PHENIX and STAR detectors are investigating polarized 
glue in the proton.
Measurements of $\Delta g/g$ from RHIC 
are sensitive to gluon polarization in the range
$0.02 < x_g < 0.3$
($\sqrt{s} = 200$ GeV) and   
$0.06 < x_g < 0.4$
($\sqrt{s} = 62.4$ GeV) 
for the neutral pion ${\cal A}_{LL}$ measured by PHENIX 
\cite{phenix200,phenix62}
and 
inclusive jet production measured by STAR at 200 GeV centre of mass energy
\cite{star200,star2008}.

The RHIC data for these asymmetries appear in Fig. 4, together with 
the expectations based on different NLO fits to inclusive $g_1$ data. 
In Fig. 4 the curves 
``GRSV-min'' (or ``$\Delta g=0$''), 
``GRSV-std'', 
``GRSV-max'' (or ``$\Delta g=g$'') and 
``$\Delta g = -g$'' 
correspond to a first moment of $\Delta g \sim 0.1$, $0.4$, $1.9$ and $-1.8$ 
respectively at $Q^2 \sim 1$ GeV$^2$ in the analysis of Ref.[37].
The data are consistent with small gluon polarization in the measured
kinematics and the value extracted 
by PHENIX from their $\sqrt{s} = 200$ GeV data is \cite{phenix200}
\begin{equation}
\Delta g_{\rm GRSV}^{[0.02,0.3]} = 0.2 \pm 0.1 (stat.) \pm 0.1 (sys.)
^{+0.0}_{-0.4} (shape) \pm 0.1 (scale)
\end{equation}
at $Q^2 = 4$ GeV$^2$.

These measurements suggest that polarized glue is,
by itself, not sufficient to resolve the difference between the small
value
of $g_A^{(0)}|_{\rm pDIS}$
and the naive constituent quark model prediction, $\sim 0.6$
through the polarized glue term 
$-3 {\alpha_s \over 2 \pi} \Delta g$. 
Note however that a gluon polarization $\sim 0.2 - 0.3$ is
would still make a significant contribution to the spin of
the proton in Eq.(10).

\subsection{Valence and Sea polarization}

\begin{figure}[b!]
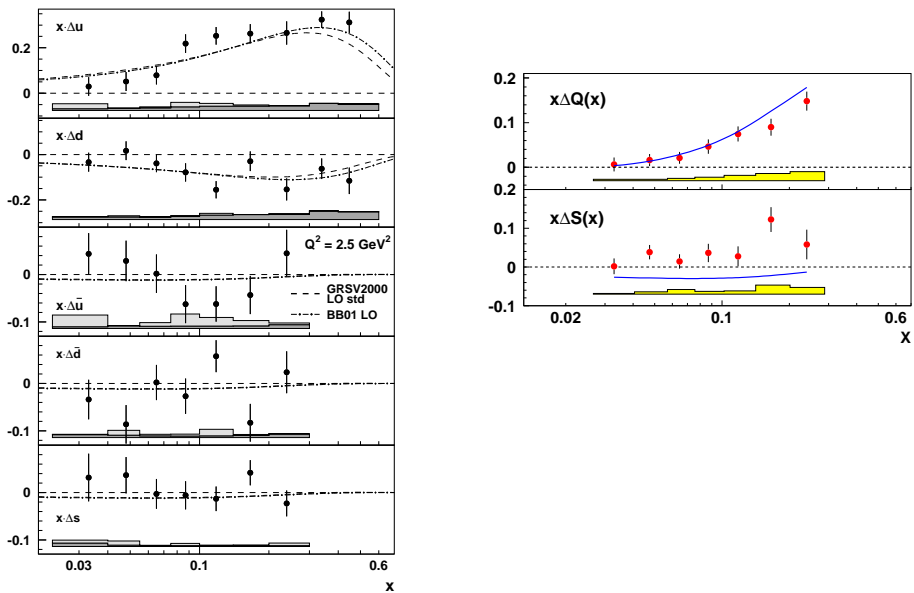

\vspace*{4.0cm} \includegraphics{hermes.epsi} \includegraphics{deltas.eps}
\vspace*{3.7cm} \caption[*]{Recent {\sc Hermes} results for the
quark and antiquark polarizations extracted from semi-inclusive DIS.
Left: (a) the flavour separation reported in Ref.[39].
Right: (b) 2008 HERMES results from charged kaon asymmetries$^{40}$.
Here $\Delta Q(x) = \Delta u(x) + \Delta d(x)$. } \label{fig:fig9}
\end{figure}

Semi-inclusive measurements of fast pions and kaons in the current
fragmentation region with final state particle identification can be
used to reconstruct the individual up, down and strange quark
contributions to the proton's spin. 
In contrast to inclusive polarized deep inelastic scattering 
where the $g_1$
structure function is deduced by detecting only the scattered
lepton, the detected particles in the semi-inclusive experiments are
high-energy (greater than 20\% of the energy of the incident photon)
charged pions and kaons in coincidence with the scattered lepton.
For large energy fraction $z=E_h/E_{\gamma} \rightarrow 1$ the most
probable occurrence is that the detected $\pi^{\pm}$ and $K^{\pm}$
contain the struck quark or antiquark in their valence Fock state.
They therefore act as a tag of the flavour of the struck quark
\cite{Close:1978}.

Figure~\ref{fig:fig9} shows the HERMES results on flavour separation 
\cite{hermesdeltas1,hermesdeltas2}.
The polarizations of the up and down quarks are positive and
negative respectively, while the sea polarization data 
are consistent with zero.
There is no evidence from this semi-inclusive data for a 
large negative strange quark
polarization. 
For the region $0.02 < x < 0.6$ 
the extracted $\Delta s$ 
integrates to the value $+0.037 \pm 0.019 \pm 0.027$\cite{hermesdeltas2}
which contrasts
with the negative value for the polarized strangeness,
Eq.(7), extracted from inclusive measurements of $g_1$. 
New COMPASS measurements\cite{windmolders} also 
show no evidence of strangeness polarization in the region $x> 0.006$.

\begin{table}[t!]
\caption{First moments for valence quark polarization
    $\Delta u_v + \Delta d_v$  
    and sea polarization $\Delta\bar{u} + \Delta\bar{d}$ from SMC$^{44}$,
HERMES$^{39}$,
and COMPASS$^{41}$.
Note that these data are dominated by the singlet cf. octet
     contributions because of the extra factor of 4 weighting in
     the flavour-decomposition of $g_1$.}
\vspace{3ex}
{\begin{tabular}{lcccr}
  Experiment  &   $x$-range & $Q^2$ (GeV$^2$) & 
{~$\Delta u_v + \Delta d_v$~} & {~$\Delta\bar{u} + \Delta\bar{d}$~} \\ 
\hline
SMC98    & 0.003--0.7 & 10  & $0.26 \pm 0.21 \pm 0.11$ & 
$0.02 \pm 0.08 \pm 0.06$ 
\\
HERMES05 & 0.023--0.6 & 2.5 & $0.43 \pm 0.07 \pm 0.06$ & 
$-0.06 \pm 0.04 \pm 0.03$ 
\\
COMPASS  & 0.006--0.7 & 10  & $0.40 \pm 0.07 \pm 0.05$  & $0.0 \pm 0.04 \pm
0.03$ \\
\end{tabular}}
\end{table}

For semi-inclusive hadron production experiments it is important 
to match the theory with the acceptance of the detector \cite{acceptance}. 
For example, the anomalous polarized gluon and low $k_t$ 
sea contributions to $g_A^{(0)}$ in Eq.~(9) have different 
transverse momentum dependence. 
The spin asymmetry for polarized $\gamma^*g$ fusion receives a 
positive contribution proportional to the quark mass squared at 
low quark $k_t^2$ plus the mass-independent negative contribution 
at large $k_t^2$ which measures the gluon polarization.
However, the magnitude of the effect 
is strongly correlated with the size of 
$\Delta g$ and vanishes for zero gluon polarization.

A direct and independent measurement of the strange quark axial-charge
through neutrino-proton elastic scattering\cite{bcsta}
would be valuable.
The axial-charge measured in $\nu p$ elastic scattering is independent of
any assumptions about the presence or absence of a subtraction at
infinity in the dispersion relation for $g_1$ and the $x \sim 0$
behaviour of $g_1$.
The W-boson production programme at RHIC\cite{rhicspin} will 
provide additional flavour-separated 
measurements of polarized up and down quarks and antiquarks.
Further measurements to push the small $x$ frontier 
would be possible with a polarized $ep$ collider\cite{trento}.

\subsection{SU(3) breaking and $g_A^{(8)}$}

\begin{table}[b!]
\caption{$g_A/g_V$ from $\beta$-decays with $F=0.46$ and $D=0.80$.}
\vspace{3ex}
{\begin{tabular}{lllr}
Process             &  measurement      &  SU(3) combination & Fit value
\\
\hline
$n \rightarrow p$     &  $1.270 \pm 0.003$  &  $F+D$   & 1.26 
\\
$\Lambda^0 \rightarrow p$ & $0.718 \pm 0.015$ & $F+{1 \over 3}D$ & 0.73 
\\
$\Sigma^- \rightarrow n$  & $-0.340 \pm 0.017$ & $F-D$ & -0.34 
\\
$\Xi^- \rightarrow \Lambda^0$ & $0.25 \pm 0.05$ & $F-{1 \over 3}D$ & 0.19 
\\
$\Xi^0 \rightarrow \Sigma^+$ & $1.21 \pm 0.05$ & $F+D$ & 1.26 
\\
\hline
\end{tabular}}
\end{table}

Given that the measured 
$\Delta s$ and $-3{\alpha_s \over 2 \pi} \Delta g$ 
contributions to $g_A^{(0)}$ are small, 
it is worthwhile to ask about the value of $g_A^{(8)}$.
The value $0.58$ 
is extracted from a 2 parameter fit to hyperon $\beta$-decays 
in terms of the SU(3) $F=0.46$ and $D=0.80$ parameters \cite{fec}
-- see Table 3.
The fit is good to 20\% accuracy \cite{leaders}.
More sophisticated fits would also include chiral corrections.
Calculations of non-singlet axial-charges in relativistic 
constituent quark models 
are sensitive 
to the confinement potential, 
effective colour-hyperfine interaction\cite{myhrer,Close:1978}, 
pion cloud 
plus additional wavefunction corrections\cite{schreiber}
chosen 
to reproduce the physical value of $g_A^{(3)}$.
These effects have the potential to reduce $g_A^{(8)}$ 
from the SU(3) value $3F-D$ to $\sim 0.5$, within the 20\% variation.
This value of $g_A^{(8)}$ 
would reduce $\Delta s_{Q^2 \rightarrow \infty}$ in Eq.(7)
to $\sim -0.05$, 
still leaving the OZI violation 
$g_A^{(0)}|_{\rm pDIS} - g_A^{(8)} \sim -0.15$ to be explained.

We have seen in Section 2 that
$g_1^d$ is flat and consistent with zero throughout the measured 
region $0.004 < x < 0.02$ where Regge extrapolation\cite{sbpvl,fec94}
would expect 
to see some divergence if there is a big contribution to $g_A^{(0)}$
from partons at small but finite $x$.
In seeking to understand the data 
we are guided by QCD anomaly theory and the special role of gluon topology.

\section{Gluon topology and the QCD axial anomaly}

In QCD one has to consider the effects of renormalization.
The flavour singlet axial vector current $J_{\mu 5}$
in Eq.(\ref{eqc53})
satisfies the anomalous divergence equation
\begin{equation}
\partial^\mu J_{\mu5}
= 6 \partial^\mu K_\mu + \sum_{i=1}^{3} 2im_i \bar{q}_i\gamma_5 q_i
\label{eqf102}
\end{equation}
where
\begin{equation}
K_{\mu} = {g^2 \over 32 \pi^2}
\epsilon_{\mu \nu \rho \sigma}
\biggl[ A^{\nu}_a \biggl( \partial^{\rho} A^{\sigma}_a
- {1 \over 3} g
f_{abc} A^{\rho}_b A^{\sigma}_c \biggr) \biggr]
\label{eqf103}
\end{equation}
is the gluonic Chern-Simons current.
Here $A^{\mu}_a$ is the gluon field and
$
\partial^{\mu} K_{\mu}
= {g^2 \over 32 \pi^2} G_{\mu \nu} {\tilde G}^{\mu \nu}
$
is the topological charge density.
Eq.(\ref{eqf102}) allows us to define a partially conserved current
$
J_{\mu 5} = J_{\mu 5}^{\rm con} + 6 K_{\mu}
$,
{\it viz.}
$
\partial^\mu J^{\rm con}_{\mu5}
= \sum_{i=1}^{3} 2im_i \bar{q}_i\gamma_5 q_i
$.

The anomaly is the physical manifestation of a clash of classical
symmetries under renormalization.
When one renormalizes the flavour-singlet axial-vector current 
operator 
the triangle diagram with one 
axial-vector current vertex and two vector current vertices is important.
One can choose an ultraviolet regularization which 
preserves current conservation (gauge-invariance) at 
the gluon vector-current vertices {\it or} one can preserve 
the partially conserved axial-vector current relation at the 
$\gamma_{\mu} \gamma_5$ vertex but not both.
Gauge invariance must win because it is dynamical and is required 
for 
renormalization leading to the anomaly on the right hand side of Eq.(12).

When we make a gauge transformation $U$
the gluon field transforms as
\begin{equation}
A_{\mu} \rightarrow U A_{\mu} U^{-1} + {i \over g} (\partial_{\mu} U) U^{-1}
\label{eqf105}
\end{equation}
and the operator $K_{\mu}$
transforms as
\begin{eqnarray}
K_{\mu} \rightarrow K_{\mu}
&+&
 i {g \over 8 \pi^2} \epsilon_{\mu \nu \alpha \beta}
\partial^{\nu}
\biggl( U^{\dagger} \partial^{\alpha} U A^{\beta} \biggr)
\nonumber \\
&+& {1 \over 24 \pi^2} \epsilon_{\mu \nu \alpha \beta}
\biggl[
(U^{\dagger} \partial^{\nu} U)
(U^{\dagger} \partial^{\alpha} U)
(U^{\dagger} \partial^{\beta} U)
\biggr]
.
\label{eqf106}
\end{eqnarray}
(Partially) conserved currents are not renormalized.
It follows that
$J_{\mu 5}^{\rm con}$
is renormalization scale invariant and the scale dependence of
$J_{\mu 5}$ associated with the factor $E(\alpha_s)$
is carried
by $K_{\mu}$.
Gauge transformations shuffle a scale invariant operator
quantity between the two operators $J_{\mu 5}^{\rm con}$
and $K_{\mu}$ whilst keeping $J_{\mu 5}$ invariant.

If we wish to understand the first moment of $g_1$ in terms of 
the matrix elements of anomalous currents
($J_{\mu 5}^{\rm con}$ and $K_{\mu}$),
then we have to understand
the forward matrix element of $K_+$ and its contribution to $g_A^{(0)}$.

Here we are fortunate in that the parton model is formulated in the
light-cone gauge ($A_+=0$) where the forward matrix elements of $K_+$
are invariant.
In the light-cone gauge the non-abelian three-gluon part of $K_+$
vanishes. The forward matrix elements of $K_+$ are then invariant
under all residual gauge degrees of freedom.
Furthermore,
in this gauge, $K_+$ measures 
the gluonic ``spin'' content of the
polarized target \cite{Jaffe:1996,Manohar:1990}.
One finds
\begin{equation}
g_A^{(0)(\rm A_+ = 0)} = \sum_q \Delta q_{\rm con}
- 3 {\alpha_s \over 2 \pi} \Delta {\cal G}
\label{eqf114}
\end{equation}
where
$\Delta q_{\rm con}$ is measured by the partially conserved
current
$J_{+5}^{\rm con}$
and
$- {\alpha_s \over 2 \pi} \Delta {\cal G}$ is measured by $K_+$.
Positive gluon polarization tends to reduce the value of
$g_A^{(0)}$
and offers a
possible source for OZI violation in $g_A^{(0)}|_{\rm inv}$.
In perturbative QCD $\Delta q_{\rm con}$ is associated with
low $k_t$ partons and is 
identified
with
$\Delta q_{\rm partons}$ and
$\Delta {\cal G}$ is identified
with $\Delta g_{\rm partons}$ (with the struck quark or antiquark
carrying $k_t^2 \sim Q^2$)
-- see Eq.(9).

If we were to work only in the light-cone gauge we might think
that we have a complete parton model description of the first
moment of $g_1$.
However, one is free to work in any gauge including a covariant
gauge where the forward matrix elements of $K_+$
are not necessarily invariant under the residual gauge degrees
of freedom \cite{Jaffe:1990a}.
Understanding the interplay between spin and
gauge invariance leads to rich and interesting physics possibilities.

For example, consider a covariant gauge.

One can show \cite{Jaffe:1990a} that the forward matrix elements of
$K_{\mu}$ are invariant under ``small'' gauge transformations
(which are topologically deformable to the identity)
but not invariant under ``large'' gauge transformations which
change the topological winding number.
Perturbative QCD involves only ``small'' gauge transformations;
``large'' gauge transformations involve strictly non-perturbative physics.
The second term on the right hand side of Eq.(\ref{eqf106})
is a total derivative;
its matrix elements vanish in the forward direction.
The third term on the right hand side of
Eq.(\ref{eqf106}) is associated with the gluon topology \cite{Cronstrom:1983}.

The topological winding number is determined by the gluonic
boundary conditions at ``infinity'',
{\it viz.}
\begin{equation}
\int d \sigma_{\mu} K^{\mu} = n 
\end{equation}
where 
$n$ is an integer and
$\sigma_{\mu}$ is 
 a large surface with boundary which is spacelike with respect
 to the positions $z_k$ of any operators or fields in the physical
 problem.
It is insensitive to local deformations of the gluon
field $A_{\mu}(z)$ or of the gauge transformation $U(z)$.
When we take the Fourier transform to momentum space
the topological structure induces a light-cone zero-mode which
can contribute to $g_1$ only at $x=0$.
Hence, we are led
to consider the possibility that there may be a
term in $g_1$ which is proportional to $\delta(x)$ \cite{bassrmp,topology}
-- hence the ${\cal C}_{\infty}$ term in Eq.(9).

Note that we are compelled to consider this possibility by the QCD 
axial anomaly and gauge invariance under large gauge transformations
-- strictly non-perturbative physics.
One can show mathematically that this contribution, if finite, 
corresponds 
to a subtraction constant in the dispersion relation for the $g_1$ 
spin structure function \cite{bassrmp}.
It is associated with the residue of 
the massless Kogut-Susskind pole that 
arises in discusion of the axial U(1) problem.
The subtraction constant, if finite, is a non-perturbative effect and 
vanishes in perturbative QCD.
It is sensitive to the mechanism of axial U(1) symmetry breaking and 
the realisation
of axial U(1) symmetry breaking by instantons:
spontaneous U(1) symmetry breaking by instantons 
naturally generates a subtraction constant whereas explicit symmetry 
breaking does not \cite{topology}.
The QCD vacuum is a Bloch superposition of states 
characterised
by non-vanishing topological winding number and non-trivial chiral 
properties.
When we put a valence quark into this vacuum it can act as a 
source which polarizes the QCD 
vacuum with net result that the spin ``dissolves'' and some 
fraction of 
the spin of the constituent quark 
is associated with non-local gluon topology with support only at 
Bjorken $x=0$.

In this scenario the finite ``$\Delta s(x)$'' 
(defined as one third 
the difference between the singlet and octet polarized quark distributions)
extracted from NLO fits to inclusive $g_1$ data corresponds, 
in part, to the area that is shifted 
to Bjorken $x=0$ through non-perturbative processes involving 
gluon topology.
(In the NLO fits to just inclusive $g_1$ data, 
 this negative ``$\Delta s(x)$'' was found to turn on strongly at 
 threshold\cite{compassnlo} in contrast to the direct measurements of
 strangeness polarization in semi-inclusive scattering.)

\section{Towards possible understanding }

Where are we in our understanding of 
the spin structure of the proton
and the small value of $g_A^{(0)}|_{\rm pDIS}$ ?
Measurements of valence, gluon and sea polarization suggest that
the polarized glue term
$-3 {\alpha_s \over 2 \pi} \Delta g_{\rm partons}$
and strange quark contribution
$\Delta s_{\rm partons}$
in Eq.(9) 
are unable to resolve the small value of $g_A^{(0)}|_{\rm pDIS}$.
The spin puzzle appears to be a property of the valence quarks.
Given that
SU(3) works well, within 20\%, 
in $\beta$-decays and the corresponding axial-charges,
then the difference between
$g_A^{(0)}|_{\rm pDIS}$ and $g_A^{(8)}$
suggests a finite subtraction 
in the $g_1$ spin dispersion relation.
If there is a finite subtraction constant,
polarized high-energy processes are not measuring the full singlet
axial-charge:
$g_A^{(0)}$ and 
the partonic contribution 
$g_A^{(0)}|_{\rm pDIS} = g_A^{(0)} - {\cal C}_{\infty}$
can be different.
Since the topological subtraction constant term affects just
the first
moment of $g_1$ and not the higher moments
it behaves like polarization at zero energy and zero momentum.
The proton spin puzzle seems to be telling us about 
the interplay of valence quarks with the complex vacuum structure of QCD.

\section*{\bf Acknowledgements}

I thank C. Aidala, A. Bravar, A. Korzenev, F. Kunne, H. Santos and 
R. Windmolders
for conversations about experimental data,
K. Aoki for help with Fig. 4, 
and 
B.L. Ioffe and A. W. Thomas for discussions about theoretical issues.
The research of
SDB is supported by the Austrian Science Fund (grant P20436).


\begin{thebibliography}{99}
%
%
\bibitem{emc}
European Muon Collab. (J. Ashman {\it et al.}), Phys. Lett. B206 (1988) 364.
%
\bibitem{bassrmp}
S. D. Bass, Rev. Mod. Phys. 77 (2005) 1257.
%
\bibitem{bassbook}
S. D. Bass,  {\it The Spin structure of the proton}
(World Scientific, 2008). 
%
\bibitem{Larin:1997}
S. A. Larin, T. van Ritbergen and J. A. M. Vermaseren
Phys. Lett. B404 (1997) 153.
%
\bibitem{PDG:2004}
Particle Data Group: C. Amsler {\it et al.}, Phys. Lett. B667 (2008) 1.
%
\bibitem{fec}
F. E. Close and R. G. Roberts, Phys. Lett. B316 (1993) 165.
%
\bibitem{compassnlo}
COMPASS Collab. (V. Yu. Alexakhin {\it et al.}), Phys. Lett. B647 (2007) 8.
%
\bibitem{Ellis:1974}
J. Ellis and R. L. Jaffe, Phys. Rev. D9 (1974) 1444.
%
\bibitem{bjsr}
Spin Muon Collab. (B. Adeva {\it et al.}), Phys. Rev. D58 (1998) 112002.
%
\bibitem{hermes}
HERMES Collab. (A. Airapetian {\it et al.}) Phys. Rev. D75 (2007) 012007.
%
\bibitem{epja}
S. D. Bass, Eur.\ Phys.\ J.\ A5 (1999) 17.
%
\bibitem{bassmb}
S. D. Bass, Mod. Phys. Lett. A22 (2007) 1005.
%
\bibitem{pvl}
J. R. Cudell, A. Donnachie and P. V. Landshoff, Phys. Lett. B448 (1999) 281.
%
\bibitem{bt91}
S. D. Bass and A. W. Thomas, J. Phys. G19 (1993) 925.
%
\bibitem{sbpvl}
S. D. Bass and P. V. Landshoff, Phys. Lett. B336 (1994) 537.
%
\bibitem{fec94}
F. E. Close and R. G. Roberts, Phys. Lett. B336 (1994) 257.
%
\bibitem{ar}
G. Altarelli and G. G. Ross, Phys. Lett. B212 (1988) 391.
%
\bibitem{et}
A. V. Efremov and O. Teryaev, JINR Report No. E2-88-287.
%
\bibitem{ccm}
R. D. Carlitz, J. C. Collins and A. Mueller, Phys. Lett. B214 (1988) 229.
%
\bibitem{bint}
S. D. Bass, B. L. Ioffe, N. N. Nikolaev and A. W. Thomas,
J. Moscow Phys. Soc. 1 (1991) 317.
%
\bibitem{topology}
S. D. Bass, Mod. Phys. Lett. A13 (1998) 791.
%
\bibitem{tgv}
S. Narison, G. M. Shore and G. Veneziano, Nucl. Phys. B433 (1995) 209.
%
\bibitem{shore}
G. M. Shore, {\tt hep-ph/0701171}.
%
\bibitem{hf}
H. Fritzsch, Phys. Lett. B229 (1989) 122.
%
\bibitem{bass99}
S. D. Bass, Phys. Lett. B463 (1999) 286.
%
\bibitem{leader}
E. Leader, A. V. Sidorov and D. B. Stamenov, Phys. Rev. D75 (2007) 074027.
%
\bibitem{Adams:1994}
FNAL E581/704 Collab. (D. L. Adams {\it et al.}), Phys. Lett. B336 (1994) 269.
%
\bibitem{compasscharm}
COMPASS Collab. (M. Alekseev {\it et al.}), {\tt arXiv:0904.3209 [hep-ex]}.
%
\bibitem{compassdeltag}
COMPASS Collab. (E. S. Ageev {\it et al.}), Phys. Lett. B633 (2006) 25.
%
\bibitem{hermesdeltag}
HERMES Collab. (A. Airapetian {\it et al.}), Phys. Rev. Lett. 84 (2000) 2584.
%
\bibitem{smcdeltag}
Spin Muon Collab. (B. Adeva {\it et al.}), Phys. Rev. D70 (2004) 012002.
%
\bibitem{Mallot:2006}
G. K. Mallot, {\tt hep-ex/0612055}.
%
\bibitem{phenix200}
PHENIX Collab. (A. Adare {\it et al.}), {\tt arXiv:0810.0694 [hep-ex]}.
%
\bibitem{phenix62}
PHENIX Collab. (A. Adare {\it et al.}), Phys. Rev. D79 (2009) 012003.
%
\bibitem{star200}
STAR Collab. (B. I. Abelev {\it et al.}), Phys. Rev. Lett. 100 (2008) 232003.
%
\bibitem{star2008}
STAR Collab.: C. A. Cagliardi, {\tt arXiv:0808.0858 [hep-ex]}.
%
\bibitem{grsv}
M. Gl\"uck, E. Reya, M. Stratmann and W. Vogelsang,
Phys. Rev. D63 (2001) 094005.
%
\bibitem{Close:1978}
F. E. Close, {\it An Introduction to Quarks and Partons}
(Academic, N.Y., 1978).
%
\bibitem{hermesdeltas1}
HERMES Collab. 
(A. Airapetian {\it et al.}), Phys. Rev. Lett. 92 (2004) 012005.
%
\bibitem{hermesdeltas2}
HERMES Collab. 
(A. Airapetian {\it et al.}), Phys. Lett. B666 (2008) 446.
%
\bibitem{Korzenev:2007}
COMPASS Collab. (M. Alekseev {\it et al.}), Phys. Lett. B660 (2008) 458.
%
\bibitem{windmolders}
COMPASS Collab.: R. Windmolders, 
{\tt arXiv:0901.3690 [hep-ex]}.
%
\bibitem{acceptance}
S. D. Bass, Phys. Rev. D67 (2003) 097502.
%
\bibitem{Adeva:1998}
SMC Collab. (B. Adeva {\it et al.}), Phys. Lett. B420 (1998) 180.
%
\bibitem{bcsta}
S. D. Bass, R. J. Crewther, F. M. Steffens and A. W. Thomas,
Phys. Rev. D66 (2002) 031901 (R).
%
\bibitem{rhicspin}
G. Bunce, N. Saito, J. Soffer and W. Vogelsang,
Ann. Rev. Nucl. Part. Sci. 50 (2000) 525.
%
\bibitem{trento}
S. D. Bass and A. De Roeck, Nucl. Phys. B (Proc. Suppl.) 105 (2002) 1.
%
\bibitem{leaders}
E. Leader and D. B. Stamenov, Phys. Rev. D67 (2003) 037503.
%
\bibitem{myhrer}
F. Myhrer and A. W. Thomas, Phys. Rev. D38 (1988) 1633.
%
\bibitem{schreiber}
A. W. Schreiber and A. W. Thomas, Phys. Lett. B215 (1988) 141.
%
\bibitem{Jaffe:1996}
R. L. Jaffe, Phys. Lett. B365 (1996) 359.
%
\bibitem{Manohar:1990}
A. V. Manohar, Phys. Rev. Lett. 65 (1990) 2511.
%
\bibitem{Jaffe:1990a}
R. L. Jaffe and A. Manohar, Nucl. Phys. B337 (1990) 509.
%
\bibitem{Cronstrom:1983}
C. Cronstr\"om and J. Mickelsson,
J. Math. Phys. 24 (1983) 2528.
%
\end{thebibliography}
\end{document}